\documentclass[12pt]{article}

\usepackage{scicite}

\usepackage{times}
\usepackage{mathtools}
\usepackage{cases}
 \usepackage{graphics,amssymb,amsmath,epsfig,color}
 \usepackage{graphicx}

\newenvironment{sciabstract}{%
\begin{quote} \bf}
{\end{quote}}

\newcounter{lastnote}
\newenvironment{scilastnote}{%
\setcounter{lastnote}{\value{enumiv}}%
\addtocounter{lastnote}{+1}%
\begin{list}%
{\arabic{lastnote}.}
{\setlength{\leftmargin}{.22in}}
{\setlength{\labelsep}{.5em}}}
{\end{list}}

\title{Mechanisms of light harvesting by photosystem II in plants}

\author
{Kapil Amarnath,$^{1\ast}$ Doran I. G. Bennett,$^{1\ast}$ Anna R. Schneider,$^{2}$\\
Graham R. Fleming$^{1\ast}$\\
\\
\normalsize{$^{1}$Department of Chemistry, University of California, Berkeley, CA}\\
\normalsize{and Physical Biosciences Division, Lawrence Berkeley National Lab, Berkeley, CA}\\
\normalsize{$^{2}$Biophysics Graduate Group, University of California, Berkeley}\\
\\
\normalsize{$^\ast$To whom correspondence should be addressed; E-mail: kapil@alum.mit.edu,}\\
\normalsize{bennett.doran@gmail.com, grfleming@lbl.gov}
}
\date{}

\begin{document} 


\baselineskip24pt


\maketitle


\begin{sciabstract}

%Abstracts explain to the general reader why the research was done and why the results are important. They should start with some brief BACKGROUND information: a sentence giving a broad introduction to the field comprehensible to the general reader, and then a sentence of more detailed background specific to your study. This should be followed by the RESULTS, or if the paper is more methods/technique oriented an explanation of OBJECTIVES/METHODS and then the RESULTS. The final sentence should outline the main CONCLUSIONS of the study, in terms that will be comprehensible to all our readers. The abstract should be 125 words or less

Light harvesting by photosystem II (PSII) in plants is highly efficient and acclimates to rapid changes in the intensity of sunlight. However, the mechanisms of PSII light harvesting have remained experimentally inaccessible. Using a structure-based model of excitation energy flow in 200 nanometer (nm) x 200 nm patches of the grana membrane, where PSII is located, we accurately simulated chlorophyll fluorescence decay data with no free parameters. Excitation movement through the light harvesting antenna is diffusive, but becomes subdiffusive in the presence of charge separation at reaction centers. The influence of membrane morphology on light harvesting efficiency is determined by the excitation diffusion length of 50 nm in the antenna. Our model provides the basis for understanding how nonphotochemical quenching mechanisms affect PSII light harvesting in grana membranes.

\end{sciabstract}

% In setting up this template for *Science* papers, we've used both
% the \section* command and the \paragraph* command for topical
% divisions.  Which you use will of course depend on the type of paper
% you're writing.  Review Articles tend to have displayed headings, for
% which \section* is more appropriate; Research Articles, when they have
% formal topical divisions at all, tend to signal them with bold text
% that runs into the paragraph, for which \paragraph* is the right
% choice.  Either way, use the asterisk (*) modifier, as shown, to

\section*{Main Text}

% 1) intro. PSII light harvesting and why we need a model.
Photosynthetic light harvesting is a multi-scale process that spans from tens of femtoseconds to minutes, and from angstroms to hundreds of nanometers\cite{Blankenship2002}. In green plants, photosynthetic light harvesting is performed by photosystems I and II, which are located in separate regions of the thylakoid membrane\cite{Dekker2005}. Photosystem II (PSII), which provides the electrons that drive photosynthesis, is located in the grana stacks of the thylakoid. PSII light harvesting starts when pigments bound to antenna proteins absorb sunlight. The nascent excitation energy is transferred to reaction centers (RC) where it is converted to chemical energy via charge separation. PSII is both an efficient and adaptable light harvesting material. At maximum light harvesting capacity, the average time for excitation capture by PSII ($\sim$300 ps\cite{vanOort2010,Holzwarth2009}) is significantly less than the excited state lifetime of pigments in the grana ($\sim$2 ns\cite{Gilmore1995,Belgio2012}), which results in a $>$80\% efficiency for converting absorbed sunlight into chemical energy\cite{Baker2008}. In addition, PSII acclimates to changes in sunlight intensity\cite{Ruban2012}, wavelength\cite{Rochaix2011}, and downstream metabolism\cite{Kanazawa2002}, which occur on the millisecond to minutes timescale. The final outcome of excitation in the grana is determined by the interplay between the rates of excitation energy transfer through the antenna and processes that irreversibly quench excitation (e.g.~fluorescence, non-radiative processes, and productive photochemistry in the RC). Time-resolved chlorophyll fluorescence, which is currently the primary technique for measuring PSII light harvesting in a variety of environmental conditions \emph{in vivo}\cite{Zaks2013}, does not have sufficient spatiotemporal resolution to fully characterize the light harvesting kinetics of PSII\cite{vanAmerongen2013}. Thus a model for PSII light harvesting that accurately captures its underlying dynamics is needed to understand the physical principles that give rise to its functionality. Such principles are essential for understanding the molecular mechanisms of photosynthesis and could be used to engineer artificial light harvesting systems with properties that mimic PSII\cite{Croce2014,Scholes2011,Fleming2012}.  \

% 2) Problems with previous models.
Current models of light harvesting in PSII treat the rates of energy transfer and trapping as parameters whose values are determined by fitting to chlorophyll fluorescence data\cite{Broess2008,Holzwarth2009,Chmeliov2015}. Conceptually, these models mainly fit within the canonical ``lake'' and ``puddle'' models\cite{Robinson1966}, which are currently used to describe the dynamics of energy transfer and trapping by PSII on the hundreds of picoseconds time scale in grana membranes. In the lake model, each excitation traverses the grana membrane sufficiently quickly to access all RCs equally. In the puddle model, excitation movement is spatially limited to the smallest photosynthetic unit, which is thought to be a photosystem II supercomplex (PSII-S) and a few surrounding major light harvesting complex II (LHCII) antenna (Fig.~1A, inset)\cite{Caffarri2011}. PSII-S consists of two reaction centers and a small number of antenna proteins\cite{Caffarri2009}. Up to now, models with free parameters for energy transfer and trapping in PSII have been considered accurate if they can fit chlorophyll fluorescence decay data well. However, multiple models that implicitly assume either the lake or puddle model can fit chlorophyll fluorescence decay data equally well\cite{Broess2008,Holzwarth2009}. As a result, it remains unclear whether the fitted rates found by these methods are physically meaningful and, more broadly, to what extent either the lake or puddle models are appropriate for describing PSII light harvesting. It is well-established that simulating the ultrafast excitation energy transfer dynamics in individual pigment-protein complexes requires a structure-based approach that correctly accounts for each chromophore\cite{Novoderezhkin2010,Renger2013b}. We previously showed that this approach is required even in a small group of pigment-protein complexes in our model of light harvesting within isolated PSII-S\cite{Bennett2013}. Here, we demonstrate that an accurate model for PSII light harvesting in grana membranes must be based on a correct description of the picosecond dynamics that occur at the nanometer scale of individual pigments. \

% 3) Our model
We constructed a parameter-less model for PSII light harvesting that is firmly based on an accurate physical description of the nanoscopic energy transfer dynamics in grana membranes. We performed Monte Carlo simulations on 200 nm x 200 nm patches of the grana membrane\cite{Schneider2013} (see Methods, below) to generate examples of the mixed (Fig.~1A) and segregated (Fig.~1B) organizations previously observed\cite{Dekker2005,Staehelin2003}. In the segregated membrane, PSII-S and LHCII separate into so-called PSII-S crystalline arrays and LHCII aggregates. The Monte Carlo model contains a small number of energetic interactions based on \emph{in vivo} phenomenology and provides plausible locations of LHCII in grana, which are difficult to visualize using existing imaging techniques\cite{Schneider2013}. We superimposed the crystal structures of the chlorophyll pigments in PSII-S \cite{Caffarri2009} and LHCII \cite{Liu2004} on these simulations to establish the locations of all of the chlorophylls in a grana patch (Fig.~1A, inset). In previous work on PSII-S, we grouped chlorophylls into highly energetically coupled clusters called domains and demonstrated that it is sufficient to describe the excitation dynamics at the domain level\cite{Bennett2013}. We assume that the kinetics within LHCII and PSII-S are the same as calculated previously on isolated complexes\cite{Bennett2013}. Inhomogeneously averaged rates of energy transfer between domains on different complexes were calculated using Generalized F{\"o}rster theory (Methods). We modeled photochemistry in the reaction centers with two phenomenological ``radical pair'' (RP) states. The rate constants between the RP states are the same as those established in previous work on the PSII-S\cite{Bennett2013}. Using our method we can routinely construct a rate matrix for grana membrane patches that contain more pigments (up to 10$^5$) than any natural light harvesting system previously modeled. \

% 4) Our model can accurately simulate the complex dynamics underlying fluorescence decay data
Our model accurately simulates chlorophyll fluorescence data and demonstrates that extracting the amplitude and lifetime components from a simple fit does not capture the complex kinetics of PSII light harvesting. The comparison of the simulation of the mixed membrane (black solid line) with experimental data from thylakoids in ref.~\cite{vanOort2010} (red dotted line) is shown in Fig.~1C. We expected the mixed membrane to be the dominant morphology of the measured thylakoids based on the conditions in which the plants (\emph{Arabidopsis thaliana}) were grown\cite{vanOort2010,Onoa2014}. The excellent agreement of the simulation with the experimental data, which was not guaranteed \emph{a priori}, suggests that our parameter-free model correctly describes excitation energy transport during PSII light harvesting. The simulated decay can be fit well to a sum of a few exponentials (Fig.~1C, inset, green bars), as is frequently done to extract the amplitude and lifetime components\cite{Zaks2013}. However, the fit does not capture the complex distribution calculated from the rate matrix (Fig.~1C, inset, black bars). Fit-based models that are only sufficiently complex to fit chlorophyll fluorescence decays well will not accurately characterize the underlying PSII light harvesting dynamics. \

\begin{figure}
\centering
\includegraphics[scale=1]{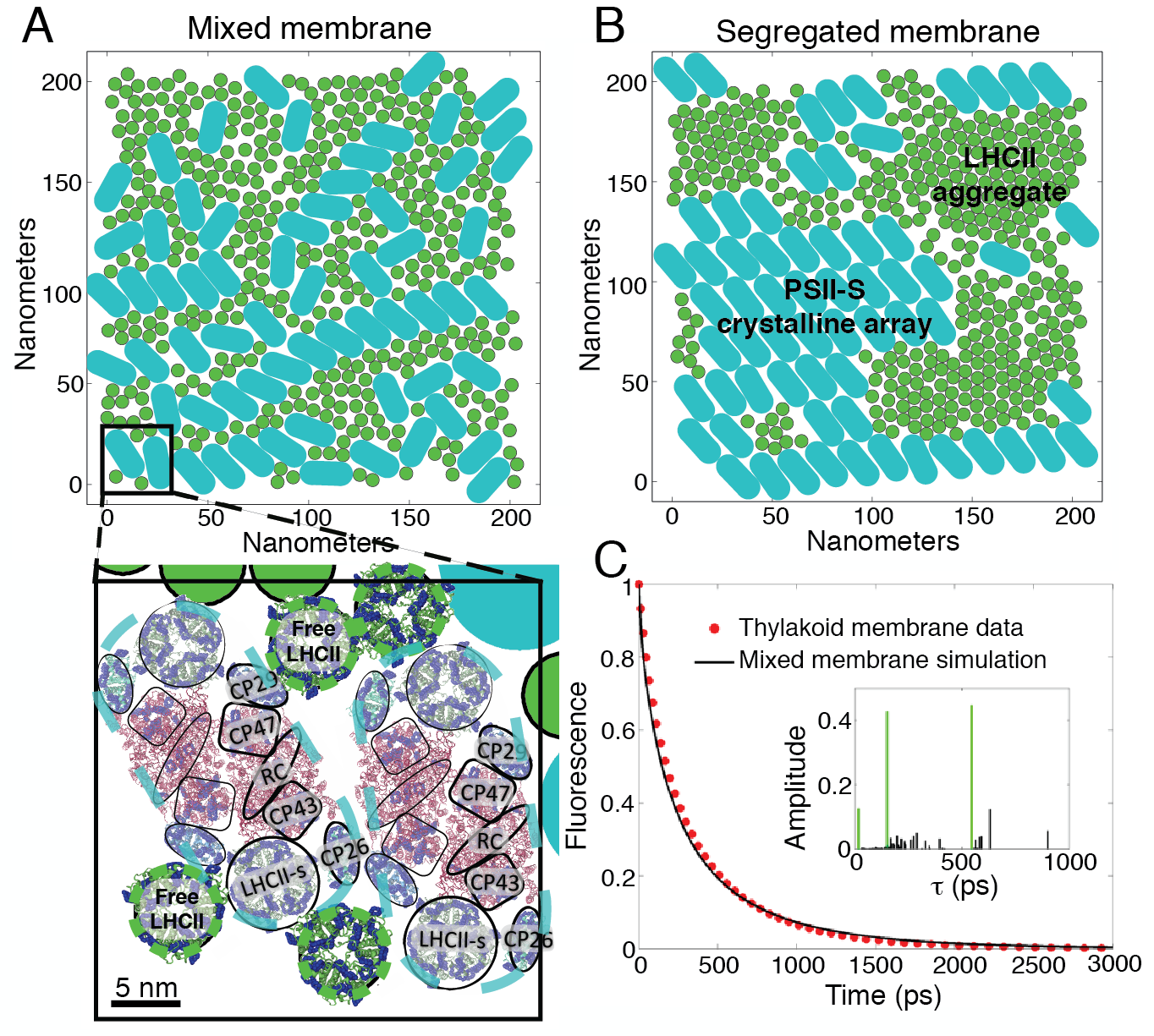}
  \caption{\textbf{Structure-based modeling of energy transfer is required to accurately simulate PSII light harvesting.} (A) and (B) The representative mixed and segregated membrane morphologies generated using Monte Carlo simulations and used throughout this work. PSII supercomplexes (PSII-S) are indicated by the teal discorectangles, while major light harvesting antenna complexes (LHCII) are indicated by the green circles. The segregated membrane forms PSII-S crystalline arrays and LHCII aggregates. As shown schematically in the inset in (A), existing crystal structures of PSII-S \cite{Caffarri2009} and LHCII \cite{Liu2004} were overlaid on these membrane patches to establish the locations of all chlorophyll pigments. The teal and green dashed lines outline the excluded area associated with PSII-S and LHCII in the Monte Carlo simulations, respectively. The chlorophyll pigments are indicated in blue, while the protein is depicted by the cartoon ribbon. PSII-S is a 2-fold symmetric dimer of pigment-protein complexes that are outlined by black lines. LHCII-s, CP26, CP29, CP43, and CP47 are antenna proteins, while RC indicates the reaction center. The inhomogeneously-averaged rates of energy transfer between strongly-coupled clusters of pigments were calculated using Generalized F{\"o}rster theory. (C) Simulated fluorescence decay of the mixed membrane (solid black line) and the PSII-component of experimental fluorescence decay data from thylakoid membranes from ref.~\cite{vanOort2010} (red, dotted line). The inset shows the lifetime components and amplitudes of the simulated decay as calculated using our model (black bars) or by fitting to three exponential decays (green bars).}
  \label{fgr:fig1}
\end{figure}

% 5) Neither lake nor puddle; need new emergent picture.
Our simulation of the spatiotemporal dynamics of chlorophyll excitation in the grana allowed us to examine the assumptions underpinning the lake and puddle models. Upon uniform initial excitation of either the mixed or segregated membranes, the lake model predicts that the excitation distribution will quickly reach a steady-state spatial distribution. However, we did not observe a steady-state distribution for either the mixed (Movie S1) or segregated (Movie S2) membranes. The puddle model predicts that there are a few (2-3) different types of trapping dynamics at reaction centers throughout the membrane. On the contrary, we observed trapping dynamics (Movies S3 and S4) that vary considerably between different PSII-S. Neither the lake nor the puddle models accurately describe excitation dynamics on the hundreds of picoseconds timescale and hundreds of nanometers length scale of the grana membrane. \

% 6) Diffusive motion 
To understand how excitation moves through the grana, we simulated excitation energy flow from single pigment-protein complexes in LHCII aggregates, PSII crystalline arrays, and mixed membranes (Movies S5-S9). To determine the effect of charge separation on excitation movement, we simulated the PSII crystalline arrays and mixed membranes both with and without the radical pair (RP) states in the reaction centers. We quantified excitation transport in these simulations by calculating the time-dependence of the variance of the excitation probability distribution (Fig.~2A). Transport across the grana was well described by fitting the equation 
\begin{equation}
\label{diff}
\sigma^2(t) - \sigma^2(0) = At^{\alpha},
\end{equation}
where $\sigma^2(t)$ is the variance at time $t$, $\sigma^2(0)$ is the variance of the initial distribution of excitation, and $A$ and $\alpha$ are fit parameters (Fig.~S4B). For $\alpha = 1$, transport is diffusive, while subdiffusive transport results if $\alpha$ is significantly less than 1. For the LHCII aggregate (Fig.~2A, green triangles) and for the mixed membranes and PSII crystalline arrays without RP states (Fig.~2A, red and blue dashed lines, respectively), $\alpha \approx 1$ and thus transport within the antenna can be considered diffusive. However, transport for both the mixed and PSII crystal cases with RP states (Fig.~2A, red and blue solid lines, respectively) is sub-diffusive ($\alpha< 0.75$). Subdiffusivity can occur when the energetic differences between sites is on the order of or greater than $k_BT$. We calculated the $\Delta G$ for the RC$\rightarrow$RP1 (radical pair 1 ) step to be -5.5$k_BT$ on the basis of our previously published rates\cite{Bennett2013}. Thus, RP1 serves as an energetic trap that causes subdiffusive transport. The energy transfer rates in our model are averaged over inhomogeneous realizations, which could mitigate the slow-down of diffusion that occurs when the width of the inhomogeneous distribution is greater than $k_BT$\cite{Akselrod2014}. The largest standard deviation of exciton energies across an inhomogeneous distribution in our model is 107 cm$^{-1}$, which is significantly less than the value of $k_BT$ at room temperature (210 cm$^{-1}$). We note that the diffusive picture of excitation energy transport elaborated here is consistent with predictions made from the PSII-S model \cite{Bennett2013} and suggests that, in the grana, excitation energy flows neither ``directionally'' nor energetically downhill on ``preferred pathways'' \cite{Croce2011}.  \

\begin{figure}
\centering
\includegraphics[scale=1]{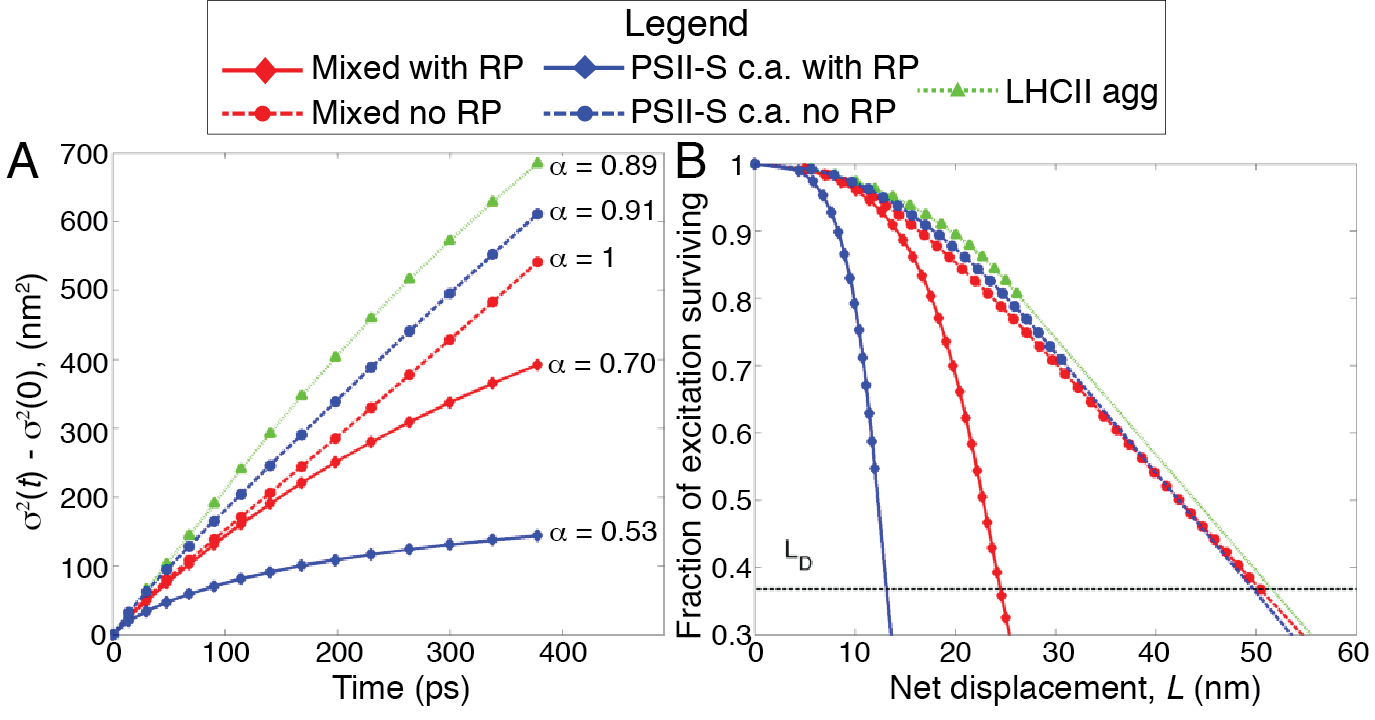}
  \caption{\textbf{Excitation transport in grana membranes.} Simulation of excitation movement in the five grana membrane configurations shown in the legend: mixed membrane with and without the radical pair (RP) states in the reaction center, PSII-S crystalline array (PSII-S c.a.) with and without the RP states, and LHCII aggregate (LHCII agg). In each case, excitation was initiated on a single pigment-protein complex. (A) The change in the spread of excitation over time. The diffusion exponent $\alpha$ (see text for details) is shown on the right of the plot. (B) Fraction of surviving excitation as a function of net displacement $L$ (eq.~2) from the initial starting point. The dashed line, where the fraction of surviving excitation is $1/e$, demarcates the excitation diffusion length ($L_D$). The dimensions of some of the configurations were too small to calculate an $L_D$, so linear extrapolation was used to approximate it (line segments that do not include markers).}
  \label{fgr:fig2}
\end{figure}

% 7) D and L_D
We calculated the excitation diffusion length, $L_D$, and the diffusivity, $D$, to reduce the dynamics observed above to single parameters that can be used to qualitatively understand light harvesting behavior in grana. 
\begin{equation}
\label{diff}
\sigma^2(t) - \sigma^2(0) = 4Dt = L^2
\end{equation}
A plot of $D$ as a function of time is shown in Fig.~S4A. In the case of subdiffusive transport, the diffusivities decrease in time as excitations are held by the RP1 trap state. The diffusion constants for the three cases ranged from $1-5\times10^{-3}$ cm$^2$/sec. These values agree with previous experimental work, which used singlet-singlet annihilation measurements of chloroplasts to suggest a lower bound of $D \approx 10^{-3}$ cm$^{-2}$/sec\cite{Swenberg1978}. The fraction of excitation remaining as a function of the net spatial displacement, $L$, is shown in Fig.~2B. $L_D$ is defined, by convention, as the minimum net displacement in one dimension achieved by 37\% of the excitation population. The $L_D$ for the mixed membrane and the PSII crystal with radical pair states were $\sim$25 nm and $\sim$15 nm, respectively, though there was some variation depending on the starting point of excitation (Fig.~S5). The $L_D$ for the LHCII aggregate, mixed membrane, and PSII crystalline array without radical pair states was $\sim$50 nm.  The values of $L_D$ and $D$ in the antenna compare favorably with the values observed thus far in organic thin films \cite{Lunt2009} and quantum dot arrays\cite{Akselrod2014}.\

% 8) Influence of diffusion on how membrane morphology effects photochemical yield
Using the excitation diffusion length ($L_D$), we explored the longstanding question of how membrane morphology influences the efficiency with which PSII converts absorbed light into chemical energy, the photochemical yield\cite{Croce2011,Dekker2005}. We calculated the photochemical yield in both the mixed and segregated cases to be 0.82 and 0.70, respectively, upon spatially uniform excitation across the membranes. The calculated value of 0.82 for the mixed membrane is in excellent agreement with the estimate of 0.83 derived from chlorophyll fluorescence yield measurements\cite{Baker2008}. Previous work suggested that the reduced light harvesting efficiency of segregated membranes was due to an inability of excitation initiated in LHCII aggregates to drive photochemistry \cite{vanOort2010}, resulting in the ``disconnection'' of LHCIIs from reaction centers. To explore this hypothesis we initiated excitation on each LHCII in both the mixed and segregated membranes and calculated the resulting photochemical yield, $\Phi_{\mathrm{LHCII}}$ (Fig.~3A).  Both membranes have a wide distribution of $\Phi_{\mathrm{LHCII}}$ (Fig.~3B). $\langle \Phi_{\mathrm{LHCII}}\rangle$ for the segregated membrane, 0.49, is significantly lower than that for the mixed membrane, 0.75. This difference can be explained most simply by the relative diffusion lengths in each membrane environment. In the mixed membrane case, nearly all LHCIIs exist within an $L_D$ (25 nm) of a PSII-S. In the segregated membrane, however, the LHCII aggregate(s) have a diameter comparable to $L_D$ (50 nm), resulting in substantial loss of excitation prior to reaching a PSII-S. Clearly, LHCIIs in the segregated membrane show reduced light harvesting function; however, the LHCIIs do not completely disconnect ($\Phi_{\mathrm{LHCII}} \approx 0$). To disconnect an LHCII aggregate from the neighboring PSII-S requires the aggregate diameter to exceed 100 nm, which is unlikely to occur given that a grana membrane disc has a diameter of $\sim$400 nm\cite{Dekker2005}. The physiological benefit of segregation remains unclear and may be explained by other aspects of photosynthesis, as has been proposed\cite{Kirchhoff2007,Kirchhoff2014}. \

\begin{figure}
\centering
\includegraphics[scale=1]{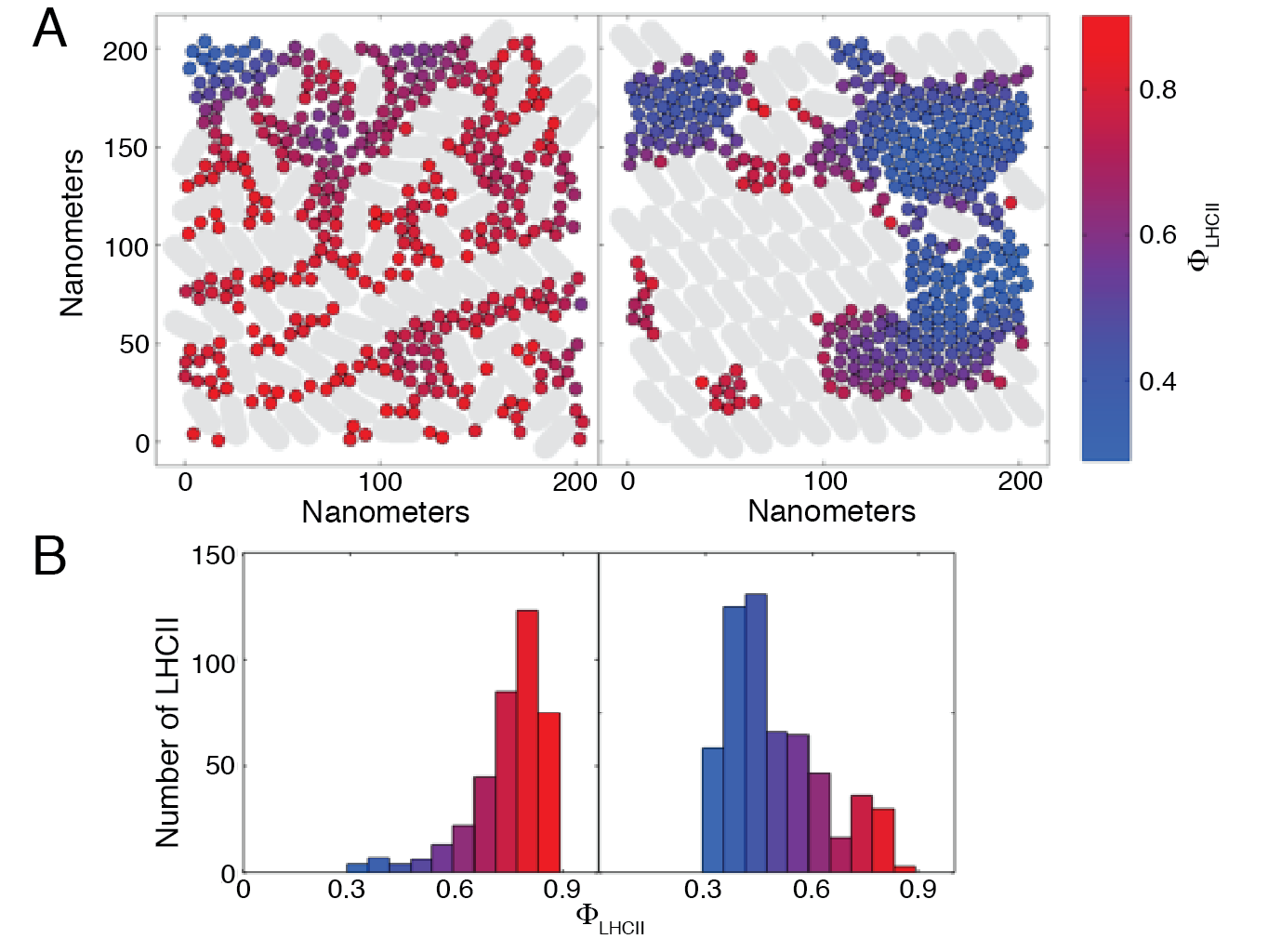}
  \caption{\textbf{Influence of grana membrane morphology on photochemical yield.} (A) Excitation was initiated at each LHCII in both the mixed (left) and segregated (right) morphologies. The color of the LHCII indicates the fraction of excitation that results in productive photochemistry ($\Phi_{\mathrm{LHCII}}$, see colorbar on far right) as simulated with our model. (B) Histograms representing the distribution of  $\Phi_{\mathrm{LHCII}}$ for the mixed and segregated membranes using the coloration from (A).}
  \label{fgr:fig3}
\end{figure}

% 9) Conclusion
Our model accurately describes the energy transfer network of the grana membrane and thus paves the way for understanding PSII light harvesting in all environmental conditions.  In intense sunlight, the rate of light absorption by the antenna exceeds the rate of trapping by reaction centers, and the need for photoprotection arises. The mechanisms underlying the nonphotochemical quenching of excess excitation in the antenna remain controversial\cite{Ruban2012}. Most of the hypothesized mechanisms of quenching are based, by necessity, on measurements performed on isolated pigment-protein complexes\cite{Zaks2013}. However, it has been unclear whether the mechanisms observed \emph{in vitro} occur \emph{in vivo}. Our model provides a unified framework for understanding to what extent picosecond dynamics observed \emph{in vitro} explain \emph{in vivo} chlorophyll fluorescence data. More broadly, PSII is connected to the rest of photosynthesis through the electron transport chain and the pH gradient across the thylakoid membrane\cite{Zhu2005,Zaks2012}. Our model could be extended to include the unappressed portion of thylakoid membranes, which contains photosystem I, and integrated into models of the entire thylakoid to fully model photosynthesis in plants. \

\section*{Methods}

\subsection*{Overview}
There has been an extensive theoretical development of energy transfer models in photosynthetic complexes composed of a few proteins. Excitation dynamics can be calculated directly from the Hamiltonian of a pigment-protein complex\cite{Ishizaki2010}. However, exact calculations of the dynamics using the Hamiltonian are computationally costly, and, while feasible on complexes with $\sim$100 pigments, are impractical for simulating dynamics in the photosystem II-containing portions of the thylakoid membrane, which contain hundreds of complexes and $>$10,000 pigments. Another common technique for calculating excitation dynamics is by perturbative treatments of the Hamiltonian, such as the Generalized F{\"o}rster and Redfield methods \cite{Novoderezhkin2010}. In photosystem II (PSII), the energetic coupling between pigments spans a wide range, such that neither method is appropriate on its own. Modified Redfield theory can interpolate between the strong and weak coupling regimes, but it does not account for dynamic localization in which coupling between the phonon modes of the protein and the pigments' excited states restrict the delocalization of excitonic states. Therefore, energy transfer in PSII has been treated using a combination of the Modified Redfield and Generalized F{\"o}rster theories \cite{Novoderezhkin2011a,Raszewski2008,Bennett2013}. Transfer within tightly coupled clusters of pigments is treated by Modified Redfield theory and transfer between these clusters is treated using Generalized F{\"o}rster theory.   \

In previous work on PSII supercomplexes (PSII-S), we used the Modified Redfield/Generalized F{\"o}rster (MR/GF) method in combination with a novel approach for defining highly coupled clusters of chlorophylls, or domains, that optimizes the separation of timescales between intra- and inter-domain transport \cite{Bennett2013}. Using these domain definitions we were able to coarse-grain the excitation dynamics at the domain level. Recent work using a more exact method (ZOFE) for calculating energy transfer dynamics has further validated our MR/GF simulations\cite{Roden2015}. Another group has performed HEOM calculations on a quadrant of the PSII-S\cite{Kreisbeck2015}. A comparison of our MR/GF simulations with these calculations gives good agreement when appropriate domain definitions are used. \

We have used the same domain definitions in our simulations of energy transfer in thylakoid membranes as were used for modeling PSII-S. We assumed that the timescales of energy transfer between domains on different complexes in the membrane are slow relative to the intra-domain timescales ($<$1 ps$^{-1}$) calculated previously\cite{Bennett2013} and used Generalized F{\"o}rster theory to calculate the rates between domains on different complexes (see Excitation energy transfer section below). The locations of all of the PSII-S and major light harvesting complex II (LHCII) antenna in a grana membrane were calculated on the basis of Monte Carlo simulations of thylakoid membranes \cite{Schneider2013} (see Monte Carlo simulations of paired grana membranes section below). The crystal structures of PSII-S \cite{Caffarri2009} and LHCII \cite{Liu2004} were overlaid on top of these simulations (see Chlorophyll configurations section below). \

The dynamics of PSII light harvesting can be represented by a kinetic network composed of 1st-order rate constants. The master equation formalism can be used to calculate the dynamics of excitation:

\begin{equation}
\dot{P}(t) = KP(t),
\end{equation}
where $K$ is a rate matrix containing the first order rate constants of excitation transfer between all compartments (which include all domains) in the network, and $P(t)$ is the vector of compartment populations \cite{Duffy2013,Zaks2013}. Within this framework, we use simple models for charge separation and trapping when excitation reaches the reaction center (see Electron transfer model section below). While there are more complex models for electron transfer within the reaction center\cite{Novoderezhkin2011b}, the two-compartment model used here was parameterized on PSII-S with different antenna sizes\cite{Bennett2013} and is sufficient to determine the overall timescales of this process. The non-radiative rate constant of decay from each domain was (2 ns)$^{-1}$\cite{Bennett2013}.  The fluorescence rate constant for each exciton was scaled by its transition dipole moment squared with the average fluorescence rate constant across all excitons set to (16 ns)$^{-1}$\cite{Bennett2013}. \

Solving eq.~1 for $P(t)$, which contains the dynamics of excitation population on all compartments in the network, gives

\begin{equation} 
P(t) = Ce^{t\Lambda}C^{-1}P(0),
\end{equation}
where $C$ is a matrix which contains the eigenvectors of $K$, $\Lambda$ is a diagonal matrix containing the eigenvalues of $K$, and $P(0)$ is the initial vector of populations. Using eq.~2, we calculate chlorophyll fluorescence decays, the dynamics of excitation over time, and the yields of the different dissipation processes available to excitation - non-radiative decay, fluorescence, and productive photochemistry in the reaction centers (see Calculations of $P(t)$ section below). 

\subsection*{Monte Carlo simulations of paired grana membranes}

Grana-scale pigment-protein complex configurations were generated via computer simulations of an extension of the model presented in ref.~\cite{Schneider2013}. Briefly, disc-shaped particles L representing LHCII complexes and rod-shaped particles P representing so-called C$_2$S$_2$ PSII-S in 2d layers $\alpha$ and $\beta$ interacted via hard-core repulsive interactions, plus the attractive energetic potentials
\begin{equation}\label{eq:mcstacking}
u_\mathrm{stack}(r_{\mathrm{L_{\alpha}-L_{\beta}}}) = 
\begin{cases}
-\epsilon_{\mathrm{stack}} \left( \frac{r_{\mathrm{L_{\alpha}-L_{\beta}}} - \sigma_{\textrm{L}}}{\sigma_{\textrm{L}}} \right) ^2 & \textrm{if $r_{\mathrm{L_{\alpha}-L_{\beta}}}<\sigma_{\textrm{L}}$} \\
0 & \textrm{otherwise,}
\end{cases}
\end{equation}
\begin{equation}\label{eq:mcmcomplex}
u_\mathrm{M}(r_{\mathrm{L_{\alpha}-P_{\alpha}}}) = 
\begin{cases}
-\epsilon_{\mathrm{M}} & \textrm{if $r_{\mathrm{L_{\alpha}-P_{\alpha}}}<\lambda_{\mathrm{M}}\sigma_{\rm L}$} \\
0 & \textrm{otherwise, and}
\end{cases}
\end{equation}
\begin{equation}\label{eq:mcagg}
u_\mathrm{agg}(r_{\mathrm{L_{\alpha}-L_{\alpha}}}) = 
\begin{cases}
-\epsilon_{\mathrm{agg}} & \textrm{if $r_{\mathrm{L_{\alpha}-L_{\alpha}}}<\lambda_{\mathrm{agg}}\sigma_{\rm L}$} \\
0 & \textrm{otherwise,}
\end{cases}
\end{equation}
with $\epsilon_{\mathrm{stack}} = 4$ $k_{\rm B}T$, $\epsilon_{\mathrm{M}} = 2$ $k_{\rm B}T$, $\lambda_{\mathrm{agg}} = 1.15$, and all other parameters as in ref.\ \cite{Schneider2013}.

The potentials in Eqs.\ \ref{eq:mcstacking} and \ref{eq:mcmcomplex} are motivated in ref.\ \cite{Schneider2013}. The square-well attraction in Equation \ref{eq:mcagg} acts a phenomenological, non-specific energetic driving force for LHCII aggregation\cite{Johnson2011}. In the ``mixed'' condition, $\epsilon_{\mathrm{agg}} = 0$ $k_{\rm B}T$, while in the ``segregated'' condition, $\epsilon_{\mathrm{agg}} = 1$ $k_{\rm B}T$.

Canonical ensemble Metropolis Monte Carlo simulations of this model were performed as in ref.\ \cite{Schneider2013}. A ratio of free LHCII to PSII-S particles of 6 was chosen to match the conditions of ref.~\cite{vanOort2010}, and a particle packing fraction of 0.75 was chosen to match typical grana conditions. Thus, 64 PSII-S particles and 384 free LHCII particles were initialized in each of two square boxes of side length 200 nm. Simulations were equilibrated with periodic boundary conditions for at least 15M Monte Carlo sweep steps.

\subsection*{Chlorophyll configurations}

Representative configurations from the stroma-side-up layers of the Monte Carlo simulations were selected for analysis. For each configuration, chlorophyll coordinates were assigned for each pigment-protein particle by aligning the center and axis of rotation of the chlorophyll coordinates from refs.~\cite{Liu2004,Caffarri2009} to the center and axis of rotation of the simulated particle (Fig.~S3). As discussed and motivated in ref.~\cite{Bennett2013}, we have substituted the structure of an LHCII monomer in the place of the minor light harvesting complexes in PSII-S. Because simulated LHCII particles were radially symmetric, an axis of rotation was randomly selected for each LHCII particle. \

In a small number of cases across the membrane, pigments belonging to one complex enter into the excluded area of a different complex. This overlap originates from a mismatch between the idealized geometries used for the Monte Carlo simulations and the real pigment structure as shown in Fig.~S2A. As a result of protein overlap, there exist a small number of analogously high transfer rates within the membrane rate matrix. In order to understand the influence of such rates on the overall description of transport in the membrane, we have removed transfer rates exceeding certain thresholds from the rate matrix and re-calculated the resulting fluorescence decay. As can be seen in Fig.~S2B, this does not alter the fluorescence decay dynamics. \

\subsection*{Excitation Energy Transfer Model}

The rate of energy transfer from a donor domain $d$ to an acceptor domain $a$, $k^{\rm{dom}}_{a \leftarrow d}$ (eq.~3), is typically calculated for each inhomogeneous realization of a pigment-protein complex using generalized F{\"o}rster theory\cite{Novoderezhkin2011a,Raszewski2008,Bennett2013}. $k^{\rm{dom}}_{a \leftarrow d}$ is the Boltzman-weighted (eq.~4) sum of the rates from the excitons in $d$ ($\vert M \rangle \in d$) to the excitons in $a$ ($\vert N \rangle \in a$). The rates between excitons are calculated using the generalized F{\"o}rster equation (eq.~5-6), where $\int_0^{\infty} \textrm{d}t A_{M}(t) F^{\ast}_{N}(t)$ is the overlap integral, $\vert V_{M, N} \vert^2$ is the electronic coupling between the two excitons, and ${U}_{\mu,M}$ is the coefficient of the $M$th exciton on the site $\mu$.  Calculation of some observable of the system, such as the fluorescence lifetime, is usually done for each realization and then averaged. However, the thylakoid membrane contains $>$10,000 chlorophylls. Generating hundreds of realizations of the population coefficients and the overlap integrals for the membrane is not only computationally intensive, but may not necessary to accurately describe the dynamics of the system at the protein length scale.
\begin{equation}
k^{\rm{dom}}_{a \leftarrow d} = \sum_{\substack{ \vert M \rangle \in d \\ \vert N \rangle \in a}} k_{ N \leftarrow M}P^{(d)}_{M} 
\end{equation}

\begin{equation}
P^{(d)}_{M} = \frac{e^{-E_M/k_BT}}{\sum_{\substack{\vert M \rangle \in d} \\ }e^{-E_M/k_BT}} 
\end{equation}

\begin{equation}
k_{M \leftarrow N} =  \frac{\vert V_{M, N} \vert^2}{\hbar^2 \pi}  \int_0^{\infty} \textrm{d}t A_{M}(t) F^{\ast}_{N}(t) 
 \end{equation}
 
\begin{equation}
\vert V_{M, N} \vert^2 = \vert \sum_{\mu,\gamma} {U}_{\mu,M}H^{\textrm{el}}_{\mu, \gamma}   {U}_{\gamma,N} \vert^2 
\end{equation}
  \
  
Using the inhomogeneously averaged rate matrix gives the correct overall dynamics and is representative of the energetics of PSII, while greatly increasing the computational efficiency. It is computationally feasible to calculate a single rate matrix for the intact thylakoid membrane using MR/GF theory through the use of supercomputers and patience. This calculation, however, is likely to be unnecessary, because a much more computationally efficient method is available for calculating the inhomogeneously averaged rate matrix, which gives the correct overall dynamics, as shown by the fluorescence lifetime comparison in Fig.~S1 for a PSII-S. While the dynamics appear to be accurate, this averaging flattens out the energy landscape and will speed up diffusion, as has been shown in quantum dot arrays\cite{Akselrod2014}. The standard deviation of the exciton energies across the ensemble of inhomogeneous realizations in PSII are significantly less than $k_BT$, suggesting that excitation movement through the grana membrane should not be greatly affected by averaging. Nonetheless, ongoing developments in making exact calculations of energy transfer more scalable \cite{Kreisbeck2014} will hopefully enable such calculations on membranes in the future.  \ 

Directly calculating the inhomogeneous average transfer rates between domains substantially reduces the computational burden because a single PSII-S contains all of the different domains that occur throughout the grana membrane. The most computationally demanding components for determining the rate matrix for a given inhomogeneous realization are the overlap integral in eq.~5 and the matrix to transform from the site to the exciton basis in eq.~6. We have assumed that domain definitions, overlap integrals, and the transformation matrices for PSII-S and LHCII are the same at the membrane level as they are in isolated complexes. As a result these terms have been computed previously for several hundred realizations of inhomogeneous broadening for the four different types of transfers (from LHCII to LHCII, PSII to PSII, LHCII to PSII and PSII to LHCII) in the membrane \cite{Bennett2013}. We tabulated these values and combined them with a calculation of the electrostatic coupling between sites (the only term that must be calculated for each domain-to-domain rate in the membrane). By using our tabulated values for the population matrices and overlap integrals, we have reduced the computational time by 3-4 orders of magnitude, turning an otherwise burdensome calculation into one that can be performed in less than 1 day on a single CPU. We note that this simplification works only because we calculate the inhomogeneously averaged rate between each domain. This allows for a much smaller sampling of the possible combinations of site energies then would be required for even a single inhomogeneous realization of the entire thylakoid membrane. \  
 
 % algorithm:
Our algorithm for calculating $\langle k^{\rm{dom}}_{a \leftarrow d} \rangle_{\mathrm{inhom.~real.}}$ for each domain pair in the membrane was as follows. We first checked if the two domains were within 60 \AA~of each other. If yes, we proceeded to calculate the average rate; if no, the rate was set equal to 0. We used 60 \AA~as a cutoff based on the distance dependence of rates in the PSII supercomplex. If the two domains were from the same complex, we used the inhomogeneously averaged rate of transfer from already performed generalized F{\"o}rster/modified Redfield calculations\cite{Bennett2013}.  If not, we noted in which type of complex the donor domain and the acceptor domain were in, either LHCII or PSII-S. We calculated $k^{\rm{dom}}_{a \leftarrow d}$ using eq.~3-6 with pre-tabulated values of the overlap integral, the population matrices, and the energies of the excitons for a few hundred inhomogeneous realizations. The electronic coupling $H^{\textrm{el}}_{\mu, \gamma}$ was uniquely calculated for that pair of domains using the ideal dipole approximation. We then averaged over these tabulated inhomogeneous realizations to get $\langle k^{\rm{dom}}_{a \leftarrow d} \rangle_{\mathrm{inhom.~real.}}$.  \

\subsection*{Electron transfer model}
Both the identity of the primary donor and the kinetics and mechanism of charge separation in the PSII reaction center remain controversial\cite{Renger2010}. The lack of agreement between the various experimental results and theory has resulted in a variety of phenomenological and conceptual models being used to interpret experimental data\cite{Broess2006,Chmeliov2015}. In previous work on PSII supercomplexes, we used the simplest kinetic model that describes the processes known to occur in the PSII reaction center\cite{Bennett2013},

\begin{equation}
\vcenter {\hbox{\includegraphics[scale=0.8]{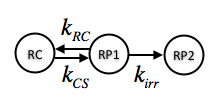}}}.
\end{equation}
Here, RC is the reaction center domain composed of the 6 pigments of the reaction center. The ``radical pair'' states RP1 and RP2 and the rate constants  $k_{RC}$, $k_{CS}$, and $k_{irr}$ are used to model the electron transfer steps in the reaction center. RP1 and RP2 are non-emissive states that do not have a direct physical analog with charge-separated states in the reaction center. Rather, this approach allows us to describe the overall process of a reversible charge separation step followed by an irreversible step and establish the approximate timescales of these events relative to energy transfer in the light harvesting antenna. The rates were previously parameterized using fluorescence decays of PSII supercomplexes of different sizes\cite{Caffarri2011}, with $\tau_{CS} = 0.64$ ps, $\tau_{RC} = 160$ ps, and $\tau_{irr} = 520$ ps ($k=1/\tau$)\cite{Bennett2013}.  

\subsection*{Calculations of $P(t)$}
Using the definitions established in the previous sections, we can write $K$ as follows:
% clarify and expand
\begin{numcases}{K_{ji}=}
k^{\mathrm{dom}}_{j \leftarrow i}, & $i,j \leq N_{\mathrm{dom}}, i\neq j$\\
k_{\mathrm{cs}}, & $i\in \mathrm{RC}, j\in \mathrm{RP1}, j>N_{\mathrm{dom}}$ \\
k_{\mathrm{rc}}, & $i\in \mathrm{RP1}, j\in \mathrm{RC}, i>N_{\mathrm{dom}}$ \\
k_{\mathrm{irr}}, & $i\in \mathrm{RP1}, j\in \mathrm{RP2},~i,j>N_{\mathrm{dom}}$\\
k_{\mathrm{dump}}, & $i \leq N_{\mathrm{dom}}, j= N_{\mathrm{comp}} - 1$\\
\tilde{k}^{fl}_{i}, & $i \leq N_{\mathrm{dom}}, j= N_{\mathrm{comp}}$\\
-\sum\limits_{k \neq i} k^{\rm{dom}}_{k \leftarrow i}, & $i=j$,
\end{numcases}
where $K_{ij}$ is a matrix element of row $i$ and column $j$, $N_{\mathrm{dom}}$ is the total number of domains, and $N_{\mathrm{comp}}$ is the total number of compartments ($N_{\mathrm{dom}}$ plus all RP1, RP2, Fl, and dump compartments). The sizes of the $K$ for the mixed and segregated membranes modeled (Fig.~1A-B, main text) were both 10882 x 10882, which meant that calculating the eigenvalues and eigenvectors required for simulating the dynamics of excitation (eq.~2) required the use of supercomputers with $\sim$30 GB of memory.  \

Fluorescence decays were calculated using the equations described in ref.~\cite{Bennett2013}. $P(0)$ in all cases was that for ChlA excitation. The photochemical yield was calculated by summing over the populations in all RP2 states at $t=\mathrm{1~sec}$. \

% 1. not sure if in vitro dynamics occurs in vivo. 
% 2. can test if in vitro dynamics plausibly explain the in vivo quenching data.

% Your references go at the end of the main text, and before the
% figures.  For this document we've used BibTeX, the .bib file
% scibib.bib, and the .bst file Science.bst.  The package scicite.sty
% was included to format the reference numbers according to *Science*
% style.

%By correctly incorporating the dynamics of energy transfer between pigments, we show that excitation energy flow through the antenna proteins in the membrane is diffusive, while reversible charge separation acts to trap excitation at RCs. Further, using the excitation diffusion length we have calculated for the different membrane environments, we explain how membrane morphology affects the quantum efficiency of light harvesting.

\bibliography{MembranePaper}

\begin{thebibliography}{10}

\bibitem{Blankenship2002}
R.~E. Blankenship, {\it {Molecular Mechanisms of Photosynthesis}\/} (Blackwell
  Science, 2002).

\bibitem{Dekker2005}
J.~P. Dekker, E.~J. Boekema, {\it Biochimica et Biophysica Acta
  (BBA)-Bioenergetics\/} {\bf 1706}, 12 (2005).

\bibitem{vanOort2010}
B.~van Oort, {\it et~al.\/}, {\it Biophysical Journal\/} {\bf 98}, 922 (2010).

\bibitem{Holzwarth2009}
A.~R. Holzwarth, Y.~Miloslavina, M.~Nilkens, P.~Jahns, {\it Chemical Physics
  Letters\/} {\bf 483}, 262 (2009).

\bibitem{Gilmore1995}
A.~M. Gilmore, T.~L. Hazlett, Govindjee, {\it Proc. Natl. Acad. Sci. U. S.
  A.\/} {\bf 92}, 2273 (1995).

\bibitem{Belgio2012}
E.~Belgio, M.~P. Johnson, S.~Juri{\'c}, A.~V. Ruban, {\it Biophysical
  journal\/} {\bf 102}, 2761 (2012).

\bibitem{Baker2008}
N.~R. Baker, {\it Annu. Rev. Plant Biol.\/} {\bf 59}, 89 (2008).

\bibitem{Ruban2012}
A.~V. Ruban, M.~P. Johnson, C.~D.~P. Duffy, {\it Biochimica et Biophysica Acta
  ({BBA}) - Bioenergetics\/} {\bf 1817}, 167 (2012).

\bibitem{Rochaix2011}
J.-D. Rochaix, {\it Biochimica et Biophysica Acta (BBA)-Bioenergetics\/} {\bf
  1807}, 375 (2011).

\bibitem{Kanazawa2002}
A.~Kanazawa, D.~M. Kramer, {\it Proceedings of the National Academy of
  Sciences\/} {\bf 99}, 12789 (2002).

\bibitem{Zaks2013}
J.~Zaks, K.~Amarnath, E.~J. Sylak-Glassman, G.~R. Fleming, {\it Photosynthesis
  Research\/} {\bf 116}, 389 (2013).

\bibitem{vanAmerongen2013}
H.~van Amerongen, R.~Croce, {\it Photosynthesis Research\/} {\bf 116}, 251
  (2013).

\bibitem{Croce2014}
R.~Croce, H.~van Amerongen, {\it Nature Chemical Biology\/} {\bf 10}, 492
  (2014).

\bibitem{Scholes2011}
G.~D. Scholes, G.~R. Fleming, A.~Olaya-Castro, R.~van Grondelle, {\it Nature
  Chemistry\/} {\bf 3}, 763 (2011).

\bibitem{Fleming2012}
G.~R. Fleming, G.~S. Schlau-Cohen, K.~Amarnath, J.~Zaks, {\it Faraday
  Discuss.\/} {\bf 155}, 27 (2012).

\bibitem{Broess2008}
K.~Broess, G.~Trinkunas, A.~van Hoek, R.~Croce, H.~van Amerongen, {\it Biochim.
  Biophys. Acta, Bioenerg.\/} {\bf 1777}, 404 (2008).

\bibitem{Chmeliov2015}
J.~Chmeliov, G.~Trinkunas, H.~van Amerongen, L.~Valkunas, {\it Photosynthesis
  Research\/} pp. in press (available at
  http://link.springer.com/article/10.1007/s11120--015--0083--3) (2015).

\bibitem{Robinson1966}
G.~W. Robinson, {\it Brookhaven Symposia in Biology\/} (1966), vol.~19, p.~16.

\bibitem{Caffarri2011}
S.~Caffarri, K.~Broess, R.~Croce, H.~van Amerongen, {\it Biophysical Journal\/}
  {\bf 100}, 2094 (2011).

\bibitem{Caffarri2009}
S.~Caffarri, R.~Kou\v{r}il, S.~Kereiche, E.~J. Boekema, R.~Croce, {\it EMBO
  J.\/} {\bf 28}, 3052 (2009).

\bibitem{Novoderezhkin2010}
V.~I. Novoderezhkin, R.~van Grondelle, {\it Physical Chemistry Chemical
  Physics\/} {\bf 12}, 7352 (2010).

\bibitem{Renger2013b}
T.~Renger, F.~M{\"u}h, {\it Physical Chemistry Chemical Physics\/} {\bf 15},
  3348 (2013).

\bibitem{Bennett2013}
D.~I.~G. Bennett, K.~Amarnath, G.~R. Fleming, {\it Journal of the American
  Chemical Society\/} {\bf 135}, 9164 (2013).

\bibitem{Schneider2013}
A.~Schneider, P.~Geissler, {\it Biophysical Journal\/} {\bf 105}, 1161 (2013).

\bibitem{SI}
See supplementary information for more details.

\bibitem{Staehelin2003}
L.~A. Staehelin, {\it Photosynthesis Research\/} {\bf 76}, 185 (2003).

\bibitem{Liu2004}
Z.~Liu, {\it et~al.\/}, {\it Nature\/} {\bf 428}, 287 (2004).

\bibitem{Onoa2014}
B.~Onoa, {\it et~al.\/}, {\it PloS one\/} {\bf 9}, e101470 (2014).

\bibitem{Akselrod2014}
G.~M. Akselrod, {\it et~al.\/}, {\it Nano Letters\/} {\bf 14}, 3556 (2014).

\bibitem{Croce2011}
R.~Croce, H.~van Amerongen, {\it Journal of Photochemistry and Photobiology B:
  Biology\/} {\bf 104}, 142 (2011).

\bibitem{Swenberg1978}
C.~Swenberg, N.~Geacintov, J.~Breton, {\it Photochemistry and Photobiology\/}
  {\bf 28}, 999 (1978).

\bibitem{Lunt2009}
R.~R. Lunt, N.~C. Giebink, A.~A. Belak, J.~B. Benziger, S.~R. Forrest, {\it
  Journal of Applied Physics\/} {\bf 105}, 053711 (2009).

\bibitem{Kirchhoff2007}
H.~Kirchhoff, {\it et~al.\/}, {\it Biochemistry\/} {\bf 46}, 11169 (2007).

\bibitem{Kirchhoff2014}
H.~Kirchhoff, {\it Philosophical Transactions of the Royal Society B:
  Biological Sciences\/} {\bf 369}, 20130225 (2014).

\bibitem{Zhu2005}
X.-G. Zhu, N.~R. Baker, D.~R. Ort, S.~P. Long, {\it et~al.\/}, {\it Planta\/}
  {\bf 223}, 114 (2005).

\bibitem{Zaks2012}
J.~Zaks, K.~Amarnath, D.~M. Kramer, K.~K. Niyogi, G.~R. Fleming, {\it
  Proceedings of the National Academy of Sciences\/} {\bf 109}, 15757 (2012).

\bibitem{Ishizaki2010}
A.~Ishizaki, T.~R. Calhoun, G.~S. Schlau-Cohen, G.~R. Fleming, {\it Physical
  Chemistry Chemical Physics\/} {\bf 12}, 7319 (2010).

\bibitem{Novoderezhkin2011a}
V.~Novoderezhkin, A.~Marin, R.~van Grondelle, {\it Physical Chemistry Chemical
  Physics\/} {\bf 13}, 17093 (2011).

\bibitem{Raszewski2008}
G.~Raszewski, T.~Renger, {\it J. Am. Chem. Soc.\/} {\bf 130}, 4431 (2008).

\bibitem{Roden2015}
J.~J. Roden, D.~I. Bennett, K.~B. Whaley, {\it arXiv preprint
  arXiv:1501.06674\/}  (2015).

\bibitem{Kreisbeck2015}
C.~Kreisbeck, A.~Aspuru-Guzik, {\it arXiv preprint arXiv:1502.02657\/}  (2015).

\bibitem{Duffy2013}
C.~D.~P. Duffy, L.~Valkunas, A.~V. Ruban, {\it Physical Chemistry Chemical
  Physics\/} {\bf 15}, 18752 (2013).

\bibitem{Novoderezhkin2011b}
V.~I. Novoderezhkin, E.~Romero, J.~P. Dekker, R.~van Grondelle, {\it
  ChemPhysChem\/} {\bf 12}, 681 (2011).

\bibitem{Johnson2011}
M.~P. Johnson, {\it et~al.\/}, {\it The Plant Cell Online\/} {\bf 23}, 1468
  (2011).

\bibitem{Kreisbeck2014}
C.~Kreisbeck, T.~Kramer, A.~Aspuru-Guzik, {\it Journal of Chemical Theory and
  Computation\/} {\bf 10}, 4045 (2014).

\bibitem{Renger2010}
T.~Renger, E.~Schlodder, {\it ChemPhysChem\/} {\bf 11}, 1141 (2010).

\bibitem{Broess2006}
K.~Broess, {\it et~al.\/}, {\it Biophysical Journal\/} {\bf 91}, 3776 (2006).

\end{thebibliography}

\bibliographystyle{Science}

% Following is a new environment, {scilastnote}, that's defined in the
% preamble and that allows authors to add a reference at the end of the
% list that's not signaled in the text; such references are used in
% *Science* for acknowledgments of funding, help, etc.

\begin{scilastnote}
\item This research used resources of the National Energy Research Scientific Computing Center, a DOE Office of Science User Facility supported by the Office of Science of the U.S. Department of Energy under Contract No. DE-AC02-05CH11231. This work was supported by the Director, Office of Science, Office of Basic Energy Sciences, of the U.S. Department of Energy under Contract DE-AC02-05CH11231 and the Division of Chemical Sciences, Geosciences and Biosciences Division, Office of Basic Energy Sciences through Grant DEAC03-76SF000098 (at Lawrence Berkeley National Labs and U.C. Berkeley).
\end{scilastnote}

% For your review copy (i.e., the file you initially send in for
% evaluation), you can use the {figure} environment and the
% \includegraphics command to stream your figures into the text, placing
% all figures at the end.  For the final, revised manuscript for
% acceptance and production, however, PostScript or other graphics
% should not be streamed into your compliled file.  Instead, set
% captions as simple paragraphs (with a \noindent tag), setting them
% off from the rest of the text with a \clearpage as shown  below, and
% submit figures as separate files according to the Art Department's
% instructions.

%\clearpage
%
%\noindent {\bf Fig. 1.} Please do not use figure environments to set
%up your figures in the final (post-peer-review) draft, do not include graphics in your
%source code, and do not cite figures in the text using \LaTeX\
%\verb+\ref+ commands.  Instead, simply refer to the figure numbers in
%the text per {\it Science\/} style, and include the list of captions at
%the end of the document, coded as ordinary paragraphs as shown in the
%\texttt{scifile.tex} template file.  Your actual figure files should
%be submitted separately.

\end{document}